\documentclass[10pt]{article}
\usepackage{amsmath}
\usepackage{amssymb}
\usepackage{natbib}

\begin{document}
\bibliographystyle{plainnat}
\setcitestyle{numbers,square}

\title{\normalsize{GRAVITATIONALLY QUANTIZED ORBITS IN THE SOLAR SYSTEM:  \\
                   COMPUTATIONS BASED ON THE GLOBAL POLYTROPIC MODEL}
                  }

\author{Vassilis Geroyannis$^1$, Florendia Valvi$^2$, Themis Dallas$^3$   \\
        $^{1}$Department of Physics, University of Patras, Greece         \\  
        $^{2}$Department of Mathematics, University of Patras, Greece     \\
        $^{3}$Department of History, Archaeology and Social Anthropology, \\
              University of Thessaly, Greece                              \\
        $^1$vgeroyan@upatras.gr, $^2$fvalvi@upatras.gr, $^3$tgd@ha.uth.gr}

\maketitle

\begin{abstract}
The so-called ``global polytropic model'' is based on the assumption of hydrostatic equilibrium for the solar system, or for a planet's system of statellites (like the jovian system), described by the Lane-Emden differential equation. A polytropic sphere of polytropic index $n$ and radius $R_1$ represents the central component $S_1$ (Sun or planet) of a polytropic configuration with further components the polytropic spherical shells $S_2$, $S_3$, ..., defined by the pairs of radii $(R_1,\,R_2)$, $(R_2,\,R_3)$, ..., respectively. $R_1,\,R_2,\,R_3,\, \dots$, are the roots of the real part $\mathrm{Re}(\theta)$ of the complex Lane-Emden function $\theta$. Each polytropic shell is assumed to be an appropriate place for a planet, or a planet's satellite, to be ``born'' and ``live''. This scenario has been studied numerically for the cases of the solar and the jovian systems. In the present paper, the Lane-Emden differential equation is solved numerically in the complex plane by using the Fortran code \texttt{DCRKF54} (modified Runge-Kutta-Fehlberg code of fourth and fifth order for solving initial value problems in the complex plane along complex paths). We include in our numerical study some trans-Neptunian objects.  
\end{abstract}

\section{Introduction}
\label{intro}
In this study, we approach the issue on ``gravitational quantization of orbits'' in the solar system, or in systems of satellites of planets, by exclusively considering laws of classical mechanics. In particular, we take as basis of our treatment the equations of hydrostatic equilibrium for a nondistorted star or planet. These equations yield the well-known Lane--Emden differential equation, which is solved in the complex plane by using the so-called ``complex plane strategy'' (Section~\ref{theLEE}), developed by the first author for numerical treatment of certain astrophysical problems (see e.g. \citep{G88}; see also \citep{GV12}).

Classical mechanics is also used by some authors for treating this issue (see e.g. in \citep{PPL08} the so-called ``vibrating membrane model''). 
On the other hand, several investigators use ideas arising within the framework(s) of Scale Relativity \citep{HSG98}; Relativity Theory regarding the finite propagation speed of gravitational interaction \citep{G07}; and quantum mechanics, like appropriate Bohr--Sommerfeld discretization (\citep{AF97}, \citep{RR98}), or Schr\"{o}dinger-type equations \citep{OMC04}. Further details are given in Section~\ref{remarks}.

\section{The Lane--Emden Equation in the Framework of the Complex-plane Strategy}
\label{theLEE}
The so-called ``complex-plane strategy'' (CPS) proposes and applies numerical integration of ``ordinary differential equations'' (ODE, ODEs) in the complex plane, either along an interval $\mathbb{I}_r \subset \mathbb{R}$ when the independent variable $r$ is real, or along a contour $\mathfrak{C} \subset \mathbb{C}$ when $r$ is complex. Integrating in $\mathbb{C}$ is necessary when the ``initial value problem'' (IVP, IVPs) under consideration is defined on ODEs: (i) suffering from singularities and/or indeterminate forms in $\mathbb{R}$, and/or (ii) involving terms that become undefined in $\mathbb{R}$ when the independent variable $r$ exceeds a particular value. 

A detailed review of CPS is given in \citep{GKAR14} (Section~3.1). 

The equations of hydrostatic equilibrium for a nondistorted star are written as 
\begin{equation}
\frac{dP}{dr} = - \, \frac{G m \rho}{r^2}, \qquad
\frac{dm}{dr} = 4 \pi r^2 \rho,
\label{heq}
\end{equation}
where $P(r)$ is the pressure, $m(r)$ the mass inside a sphere of radius $r$, and $\rho(r)$ the density. For the polytropic models we use the ``polytropic EOS'' (\citep{Cha39}, Chapter~IV, Eq.~(1))
\begin{equation}
P = K \rho^{\,\Gamma} = K \rho^{1+\left(1/n\right)},
\label{Ppol}
\end{equation}
and the ``normalization equations'' (\citep{Cha39}, Chapter~IV, Equations~(8a), (10a), respectively) 
\begin{equation}
\rho = \lambda \, \theta^n, \qquad r = \alpha \, \xi,
\label{Rpol}
\end{equation}
where $K$ is the polytropic constant, $\Gamma$ the adiabatic index defined by $\Gamma = 1 + (1/n)$, $n \in [0,\,5) \subset \mathbb{R}$ the polytropic index, $\lambda$ the polytropic unit of density set equal to the central density of the star, $\lambda = \rho(r=0)$, and $\alpha$ the polytropic unit of length set equal to (\citep{Cha39}, Chapter~IV, Eq.~(10b))   
\begin{equation}
\alpha = \left[ \frac{(1+n) K \lambda^{\left(1/n\right) - 1}}{4 \, \pi \, G} \right]^{1/2}.
\label{apol}
\end{equation}
Thus $\theta^n$ is the density measured in such units, so-called ``classical polytropic units'' (abbreviated ``cpu''), and $\xi$ the length measured also in cpu.

By introducing Equations~(\ref{Ppol})--(\ref{Rpol}) into Equations~(\ref{heq}a, b), we obtain the so-called ``Lane--Emden equation'' (cf. \citep{Cha39}, Chapter~IV, Eq.~(11))
\begin{equation}
\frac{d^2\theta}{d\xi} + \frac{2}{\xi} \, \frac{d\theta}{d\xi} = 
                                                               - \, \theta^n,
\label{LEE}
\end{equation}
which, integrated along a specified integration interval 
\begin{equation}
\mathbb{I}_\xi = [\xi_\mathrm{start} = 0, \,\, \xi_\mathrm{end}] \subset \mathbb{R}
\label{LEEii}
\end{equation}
with initial conditions
\begin{equation}
\theta(\xi_\mathrm{start}) = 1, \qquad \theta'(\xi_\mathrm{start}) = 0
\label{ivs}
\end{equation}
yields as solution the ``Lane--Emden function'' $\theta[\mathbb{I}_\xi \subset \mathbb{R}] \subset \mathbb{R}$. 

However, the Lane--Emden equation~\eqref{LEE} involves (i) the indeterminate form $\theta'/\xi$ at the origin, and (ii) the ``raised-to-real-power'' term $\theta^n$ which becomes undefined for $\theta(\xi) < 0$ and, also, suffers from a ``non-monodromy syndrome'' in the sense that multiple-valued logarithmic functions are involved in the representation of $\theta^n$ (see e.g. \cite{Chu60}, Secs.~26--28). To avoid such syndroms, we apply to the IVP established on the equations~\eqref{LEE} and \eqref{ivs} the complex-plane strategy. In particular, we assume that the independent variable $\xi$ is a ``complex distance'', $\xi = \bar{\xi} + i \, \breve{\xi} \in \mathbb{C}$, and the integration proceeds along a complex path parallel to the real axis and at a relatively small imaginary distance from it. Hence, we perform numerical integration along a contour $\mathfrak{C} \subset \mathbb{C}$, being parallel to the real axis $\mathbb{R}$ and distancing $i\,\breve{\xi}_0$ from it, i.e. along the straight line-segment
\begin{equation}
 \mathfrak{C} =
  \bigl\{
    \xi_0 = \bar{\xi}_0 + i \, \breve{\xi}_0 \,\, \longrightarrow \,\, 
    \xi_\mathrm{end} = \bar{\xi}_\mathrm{end} + i \, \breve{\xi}_0  \bigr\},
\label{Pcontour}
\end{equation}
joining the points $\xi_0$ and $\xi_\mathrm{end}$ in $\mathbb{C}$. For the constant imaginary part $\breve{\xi}_0$ of the complex distance $\xi$, we usually take it to lie in the interval $\left[10^{-9},\, 10^{-3}\right]$.
Thus the Lane--Emden function $\theta$ becomes complex-valued function in one complex variable,
\begin{equation}
\theta[\frak{C} \subset \mathbb{C}] = \bar{\theta}[\frak{C}] +
                                      \breve{\theta}[\frak{C}].
\label{Ctheta}
\end{equation}

In the franework of CPS, the initial conditions~(\ref{ivs}) are written as
\begin{equation}
\theta(\xi_0) = \bar{\theta}_0 + 
                i \, \breve{\theta}_0, \qquad \theta'(\xi_0) = 0,
\label{ivscomplex}
\end{equation} 
where $\bar{\theta}_0=1$ ( cf. Equation~(\ref{ivs}a)). The initial value for the imaginary part $\breve{\theta}_0$ is selected to be small compared to the initial value for the real part, lying usually in the interval $[10^{-9},\,10^{-3}]$. In certain cases, this initial value can be set equal to zero, though systematic numerical experiments show that the presence of a nonzero initial value stabilizes and accelerates the complex-plane integration procedure.

Readers interested in issues of this section can find full details in \citep{GKAR14} (Section~3.2).

\section{The Global Polytropic Model: Application to the Solar and Jovian Systems}
\label{polSJS}
In the so-called ``global polytropic model'' for the solar system (\cite{Ger93}, Section~1), the primary assumption is hydrostatic equilibrium (Equation~(\ref{heq})). Due to the fact that $\theta[\frak{C} \subset \mathbb{C}] \subset \mathbb{C}$ (Equation~(\ref{Ctheta})), the real part $\bar{\theta}$ has a first root at $\xi_1 = \bar{\xi_1} + i \, \breve{\xi_0}$, as well as further roots: a second root at $\xi_2 = \bar{\xi_2} + i \, \breve{\xi_0}$, with $\bar{\xi_2} > \bar{\xi_1}$, a third root at $\xi_3 = \bar{\xi_3} + i \, \breve{\xi_0}$, with $\bar{\xi_3} > \bar{\xi_2}$, etc. Thus the polytropic sphere of polytropic index $n$ and radius $\bar{\xi}_1$ is the central component or central body $S_1$ of a ``resultant polytropic configuration'' of which further components are the polytropic spherical shells $S_2$, $S_3$, $S_4$, \dots, defined by the pairs of radii $(\bar{\xi}_1, \, \bar{\xi}_2)$, $(\bar{\xi}_2, \, \bar{\xi}_3)$, $(\bar{\xi}_3, \, \bar{\xi}_4)$, \dots, respectively. 

Each polytropic shell can be considered as an appropriate place for a planet, or a satellite, to be born and live. We speak for a planet when the central body $S_1$ simulates the Sun \cite{Ger93}; in this case, the resultant polytropic configuration represents the solar system. On the other hand, we speak for a satellite when $S_1$ simulates a planet, say the Jupiter (for the Jupiter's system of satellites see \cite{GV94}); then the resultant polytropic configutation represents the jovian system. 

The most appropriate location for a planet, or for a planet's satellite, to settle inside a polytropic shell $S_j$ is the place $\bar{\Xi}_j$ at which $|\bar{\theta}|$ takes its maximum value inside $S_j$, $\mathrm{max}|\bar{\theta}[S_j]| = |\bar{\theta}(\bar{\Xi}_j + i \, \breve{\xi}_0)|$. Now, concerning the solar system, the distance $A_\mathrm{E}$ of the Earth (the most massive inner planet) from the Sun is 
\begin{equation}
A_\mathrm{E} = 1 \, \mathrm{AU} = 214.9487 \, R_\odot;
\label{AEarth}
\end{equation}
the distance $A_\mathrm{J}$ of the Jupiter (the most massive outer planet) from the Sun is 
\begin{equation}
A_\mathrm{J} = 5.202 \, \mathrm{AU} = 1118.1632 \, R_\odot;
\label{AJup}
\end{equation}
and the distance $A_\mathrm{N}$ of the Neptune (the most massive among the most distant outer planets) from the Sun is
\begin{equation}
A_\mathrm{N} = 30.057 \, \mathrm{AU} = 6460.7132 \, R_\odot;
\label{ANept}
\end{equation}
The triplet $A_\mathrm{E}$, $A_\mathrm{J}$, $A_\mathrm{N}$ seems to be a representative triplet of distances for the solar system. 
To compute an ``optimum polytropic index'' $n_\odot$ for the Sun, we search for a particular value of the polytropic index which generates a sequence of maximum values $\mathrm{max}|\bar{\theta}[S_j]|$, $j=2,\,3,\,\dots,\,L$, with the integer $L$ taken sufficiently large, occuring at distances $\bar{\Xi}_j$, $j=2,\,3,\,\dots,\,L$ among which there are values $\bar{\Xi}_\mathrm{E}$, $\bar{\Xi}_\mathrm{J}$, $\bar{\Xi}_\mathrm{N}$, yielding respective distances $\alpha_\mathrm{E}=\alpha_\odot \, \bar{\Xi}_\mathrm{E}$, $\alpha_\mathrm{J}=\alpha_\odot \, \bar{\Xi}_\mathrm{J}$, $\alpha_\mathrm{N}=\alpha_\odot \, \bar{\Xi}_\mathrm{N}$, being as close to the given distances  $A_\mathrm{E}$, $A_\mathrm{J}$, $A_\mathrm{N}$, as possible; in agreement with Equation~(\ref{Rpol}b), the unit of length for the Sun is given  by $\alpha_\odot = R_\odot / \bar{\xi}_1$. Note that, in such computations, the astronomical unit remains invariant, equal to $214.9487 \, R_\odot$ (Equation~(\ref{AEarth})) irrespective of the particular value $\bar{\xi}_1$.
Computations presented in \cite{Ger93} (Section~2, Table I) give as optimum polytropic index for the Sun the value $n_\odot = 3.23$.

In \cite{GV94} (Section~1), to find an optimum polytropic index for the Jupiter, we have used a general algorithm called A[n]. In detail, to compute an optimum polytropic index $n_\mathrm{opt}$ so that a triplet of planets (or satellites), distancing $A_{P1} < A_{P2} < A_{P3}$ from the central body, be accomodated inside the resultant polytropic configuration, we work as follows. \\
A[n]--1.~For a sequence of values $n_i$, $i=1,\,2,\,\dots,\,N_\mathrm{n}$, we compute the corresponding sequence of distances $\alpha_{pj}(n_i)=\bar{\Xi}_j(n_i)$, $j=2,\,3,\,\dots,\,L$, at which planets/satellites can be accomodated, with the integer $L$ taken sufficiently large. \\
A[n]--2.~For each sequence $\{\alpha_{pj}(n_i)\}$, we compute the two-dimensional ``array of distance ratios''
\begin{equation}
D(n_i;\,j,\,k) = \, \frac{\alpha_{pj}(n_i)}{\alpha_{pk}(n_i)}, \qquad 
j = 2,\,3,\,\dots,\,L, \qquad k = 2,\,3,\,\dots,\,L,
\label{Darray}
\end{equation} 
A[n]--3.~We scan the $N_\mathrm{n}$ arrays $D(n_i;\,j,\,k)$ in order to find a value $n_\mathrm{opt}$ generating a ``maximum number of proper levels'' related to the given ratios $R_{12} = A_{P1}/A_{P2}$ and $R_{13} = A_{P1}/A_{P3}$. Note that two particular elements $D(n_i;\,q,\,s)$ and $D(n_i;\,q,\,t)$, $t > s$, constitute a ``proper level'' (and thus increase by one the number of the proper levels in favor of the polytropic index $n_i$) if             
\begin{equation}
100 \times \frac{|R_{12} - D(n_i;\,q,\,s)|}{R_{12}} \, \leq \tau,
\end{equation}
and
\begin{equation}
100 \times \frac{|R_{13} - D(n_i;\,q,\,t)|}{R_{13}} \, \leq \tau,
\end{equation}
verified within a percent tolerance $\tau$ specified by the user.

Applying A[n] to the Jupiter's system of satellites (\cite{GV94}, Section~2, Tables~I--III), we have computed an optimal value $n_\mathrm{J} = 2.45$ for this planet.

\section{The Computations}
\label{computations}
To compile our programs, we use the gfortran compiler, licensed under the GNU General Public License (GPL; http://www.gnu.org/licenses/gpl.html). gfortran is the name of the GNU Fortran compiler belonging to the GNU Compiler Collection (GCC; http://gcc.gnu.org/). In our computer, it has been installed by the TDM-GCC ``Compiler Suite for Windows'' (http://tdm-gcc.tdragon.net/), which is free software distributed under the terms of the GPL.

Subroutines required for standard numerical procedures (e.g. interpolations of functions, rootfinding of algebraic equations, localizing extrema of functions, etc.) are taken from the SLATEC Common Mathematical Library, which is an extensive public-domain Fortran Source Code Library, incorporating several public-domain packages. The full SLATEC release is available in http://netlib.cs.utk. edu/.

To solve the complex IVPs involved in this investigation, we use the code \texttt{DCRKF54} included in the Fortran package dcrkf54.f95 \cite{GV12}. \texttt{DCRKF54} is a Runge--Kutta--Fehlberg code of fourth and fifth order modified for the purpose of solving complex IVPs, which are allowed to have high complexity in the definition of their ODEs, along contours (not necessarily simple and/or closed) prescribed as continuous chains of straight-line segments; interested readers can find full details on dcrkf54.f95 in \cite{GV12}.

The header of \texttt{DCRKF54} is given in \cite{GV12} (Section~2.1, Part~\#[000]). 
On entry to \texttt{DCRKF54}, the input arguments are assigned the values $\texttt{A}=r_\mathrm{start}$, $\texttt{B}=r_\mathrm{end}$, $\texttt{N}=n$, $\texttt{Y}=\mathbf{y}_\mathrm{start}$; \texttt{DEQS} is the subroutine computing the vector derivative(s) function $\mathbf{f}$ (\cite{GV12}, Section~2.1, Equation~(1a), and discussion preceding this equation), \texttt{HIN} an initial stepsize, \texttt{HMIN} a minimum stepsize, \texttt{HMAX} a maximum stepsize. In this work, the input parameters are assigned the values $\texttt{H}=10^{-3}$, $\texttt{HMIN}=10^{-6}$, $\texttt{HMAX}=10^{-1}$. Furthermore, the input values next to \texttt{HMAX} (discussed in \cite{GV12}, Appendix A) are assigned the values $\texttt{ATOL}=10^{-24}$, $\texttt{RTOL}=10^{-14}$ for double precision (\texttt{KD=8}, where \texttt{KD} is the overall ``kind type parameter'' explained in \cite{GV12}, Sec~2.2, discussion following Part~\#[060]), $\texttt{RTOL}=10^{-15}$ for high precision (\texttt{KD=10}), and $\texttt{QLBD}=7.5 \times 10^{-1}$. In fact, we almost use a pure relative error control, since \texttt{ATOL} is $\sim\!10$ orders of magnitude less than \texttt{RTOL}. Concerning the equality tolerance \texttt{XTOL} (\cite{GV12}, Section~2.2, Part~\#[050]), it takes the value $\texttt{XTOL}= 32 \times \texttt{EMR}$, where \texttt{EMR} is the well-known ``machine roundoff error'' (i.e. the larger real number which does not change unity when added to it). Alternatively, the user can add \texttt{XTOL} to the call sequence by modifying the header of \texttt{DCRKF54} and specify its input value in the calling program. On exit to \texttt{DCRKF54}, \texttt{A} has been hopefully advanced to \texttt{B}, \texttt{Y} is the solution vector at \texttt{B}, and \texttt{H} is the stepsize adapted so far, situation verified  by the return value $\texttt{NFLAG}=1$. If \texttt{A} needs further steps to arrive at \texttt{B}, then the return value is $\texttt{NFLAG}=2$, whence we call again \texttt{DCRKF54} leaving all of its arguments unchanged. 

In this study, integrations proceed along the following members of the contour class $\mathfrak{C_\mathrm{2Form}}$ (\cite{GV12}, Section~5, Equation~(17)) 
\begin{equation}
\mathfrak{C}_{2B} = \{(10^{-4},\,10^{-4}) \rightarrow (2.0 \times 10^2,\,10^{-4})
                    \rightarrow (10^{-4},\,10^{-4}) \},
\label{C2B}
\end{equation}
\begin{equation}
\mathfrak{C}_{2C} = \{(10^{-4},\,10^{-4}) \rightarrow (5.7 \times 10^4,\,10^{-4})
                    \rightarrow (10^{-4},\,10^{-4}) \},
\label{C2C}
\end{equation}
\begin{equation}
\mathfrak{C}_{2D} = \{(10^{-4},\,10^{-4}) \rightarrow (1.0 \times 10^7,\,10^{-4})
                    \rightarrow (10^{-4},\,10^{-4}) \},
\label{C2D}
\end{equation}
for satellites of the jovian system, planets of the solar system, and ``trans-Neptunian objects'' (TNO, TNOs), respectively. The contour class $\mathfrak{C_\mathrm{2Form}}$ represents forward-and-then-backward straight-line routes parallel and close to $\mathbb{R}$ obeying the special form~(\ref{Pcontour}). The endpoint of any such contour coincides with its start point, $\xi_\mathrm{end} = \xi_0$. Hence, the true value $\bar{\theta}_\mathrm{end}$ at the endpoint $\xi_\mathrm{end}$ coincides with its initial value $\bar{\theta}_0$ (Equation~\eqref{ivscomplex}) at the start point $\xi_0$. From the numerical analysis point of view, this is an important fact, since the global percentage errors $\%\mathcal{E}(\bar{\theta})$ owing to \texttt{DCRKF54} can be readily calculated. A discussion on various contours and their characteristics can be found in \cite{GV12} (Section~5).

\section{Numerical Results and Discussion}
The following two subsections contain numerical results for the purpose of testing the code \texttt{DCRKF54} and of comparing with previous (published) corresponding results. The third subsection contains results regarding some trans-Neptunian objects. Some earlier (unpublished) corresponding numerical results will not be quoted here, since the computations of the present study are more accurate and reliable.

\subsection{Satellites of the jovian system}
The jovian system of satellites constitutes a short-distance integration problem, since the related complex IVP is solved along the contour $\mathfrak{C}_{2B}$ (Equation~\eqref{C2B}). Integrating by the code \texttt{DCRKF54} has been proved very accurate, since the global percentage error has been found to be $\%\mathcal{E}(\bar{\theta}) \leq 2 \times 10^{-9}$; whence, the numerical results in Table~\ref{jupiter} can be safely quoted (as they do) with seven decimal digits. As said in Section~\ref{polSJS}, the optimum polytropic index for the Jupiter is $n_\mathrm{opt}(\mathrm{J}) = n_\mathrm{J} = 2.45$. All symbols involved in Table~\ref{jupiter} are explained in \cite{GV94} (see especially Section~2; comparisons can be made with respective results of Table III). In both studies, Europa, Ganymede, and Callisto occupy Shells~No~4, No~5, and No~7, respectively. 

\begin{table}
\begin{center}
\caption{The jovian system: Quantities describing the central body $S_1$, i.e. the Jupiter, and the polytropic spherical shells $S_4=(\bar{\xi}_3,\,\bar{\xi}_4)$, $S_{5}=(\bar{\xi}_{4},\,\bar{\xi}_{5})$, $S_{7}=(\bar{\xi}_{6},\,\bar{\xi}_{7})$ (Section~\ref{polSJS}) of the satellites Europa (E), Ganymede (G), and Callisto (C), respectively, computed by the code \texttt{DCRKF54(KD=10)}.\label{jupiter}}
\begin{tabular}{lr} 
\hline \hline
& \texttt{DCRKF54(KD=10)} \\
\hline 
Jupiter--Shell No                                                 & 1                  \\
\hline 
$n_\mathrm{J}$                                                    & 2.45               \\
$\bar{\xi}_1$                                                      & 5.2361414(+00)     \\
$R_\mathrm{J}$ (cm)                                                & 6.9173000(+04)     \\
$M_\mathrm{J}$ (g)                                                 & 1.8990000(+30)     \\
$\alpha_\mathrm{J}$ (cm)                                           & 1.3211000(+09)     \\
$\lambda_\mathrm{J}$ ($\mathrm{g\,cm^{-3}}$)                       & 2.9701122(+01)      \\
\hline 
Europa--Shell No                                                   & 4                  \\
\hline 
Inner radius, $\bar{\xi}_3$                                        & 3.0394242(+01)     \\
Inner radius measured in Jupiter's radii                           & 5.8047022(+00)     \\
Outer radius, $\bar{\xi}_4$                                        & 5.4281234(+01)     \\
Outer radius measured in Jupiter's radii                           & 1.0366648(+01)     \\
Radius $\alpha_\mathrm{E}$ of $\mathrm{max}|\bar{\theta}|$ measured in Jupiter's radii         
& 7.6774244(+00)     \\
Percentage error in $\alpha_\mathrm{E}$, given that $A_\mathrm{E}=9.405\,R_\mathrm{J}$   
& 18.37              \\ 
\hline 
Ganymede--Shell No                                                 & 5                  \\
\hline 
Inner radius, $\bar{\xi}_4$                                        & 5.4281234(+01)     \\
Inner radius measured in Jupiter's radii                           & 1.0366648(+01)     \\
Outer radius, $\bar{\xi}_5$                                        & 8.0249546(+01)     \\
Outer radius measured in Jupiter's radii $\bar{\xi}_1$             & 1.5326085(+01)     \\
Radius $\alpha_\mathrm{G}$ of $\mathrm{max}|\bar{\theta}|$ measured in Jupiter's radii 
& 1.3445027(+01)     \\
Percentage error in $\alpha_\mathrm{G}$, given that $A_\mathrm{G}=15.003\,R_\mathrm{J}$     
& 10.38              \\ 
\hline  
Callisto--Shell No                                                 & 7                  \\
\hline 
Inner radius, $\bar{\xi}_6$                                        & 1.1618092(+02)     \\
Inner radius measured in Jupiter's radii                           & 2.2188270(+01)     \\
Outer radius, $\bar{\xi}_7$                                        & 1.6904862(+02)     \\
Outer radius measured in Jupiter's radii                           & 3.2284961(+01)     \\
Radius $\alpha_\mathrm{C}$ of $\mathrm{max}|\bar{\theta}|$ measured in Jupiter's radii          
& 2.6775448(+01)     \\
Percentage error in $\alpha_\mathrm{C}$, given that $A_\mathrm{C}=26.388\,R_\mathrm{J}$       
& 1.47               \\   
\hline 
\end{tabular}
\end{center}
\end{table}

\subsection{Planets of the solar system}
Treating planets of the solar system is a long-distance integration problem, since the so-defined complex IVP is solved along the contour $\mathfrak{C}_{2C}$ (Equation~\eqref{C2C}). Numerical integration by the code \texttt{DCRKF54} gives very accurate results, since $\%\mathcal{E}(\bar{\theta}) \leq 2 \times 10^{-9}$; accordingly, the numerical results in Table~\ref{sun} are quoted again with seven decimal figures. As said in Section~\ref{polSJS}, the optimum polytropic index for the Sun is $n_\odot = 3.23$. All symbols involved in Table~\ref{sun} are explained in \cite{GV94} (especially in Section~2; comparisons can be made with respective results of Tables I and II). In both studies, Earth, Jupiter, and Neptune occupy Shells~No~7, No~11, and No~18, respectively.

\begin{table}
\begin{center}
\caption{The solar system: Quantities describing the central body $S_1$, i.e. the Sun, and the polytropic spherical shells $S_7=(\bar{\xi}_6,\,\bar{\xi}_7)$, $S_{11}=(\bar{\xi}_{10},\,\bar{\xi}_{11})$, $S_{18}=(\bar{\xi}_{17},\,\bar{\xi}_{18})$ of the planets Earth (E), Jupiter (J), and Neptune (N), respectively, computed by the code \texttt{DCRKF54(KD=10)}.\label{sun}}
\begin{tabular}{lr} 
\hline \hline
& \texttt{DCRKF54(KD=10)} \\
\hline 
Sun--Shell No                                                      & 1                  \\
\hline 
$n_\odot$                                                          & 3.23               \\
$\bar{\xi}_1$                                                      & 7.9169049(+00)     \\
$R_\odot$ (cm)                                                     & 6.9598000(+10)     \\
$M_\odot$ (g)                                                      & 1.9890000(+33)     \\
$\alpha_\odot$ (cm)                                                & 8.7911000(+09)     \\
$\lambda_\odot$ ($\mathrm{g\,cm^{-3}}$)                            & 1.1917082(+02)     \\
$\mathrm{AU}$ measured in $\alpha_\odot$                           & 1.7017284(+03)     \\
\hline 
Earth--Shell No                                                    & 7                  \\
\hline 
Inner radius, $\bar{\xi}_6$                                        & 1.4338943(+03)     \\
Inner radius in AU                                                 & $8.4261056(-01)$   \\
Outer radius, $\bar{\xi}_7$                                        & 2.3786878(+03)     \\
Outer radius in AU                                                 & 1.3978070(+00)     \\
Radius $\alpha_\mathrm{E}$ of $\mathrm{max}|\bar{\theta}|$ in AU   & 1.0865424(+00)     \\
Percentage error in $\alpha_\mathrm{E}$                            & 8.65               \\ 
\hline 
Jupiter--Shell No                                                  & 11                 \\
\hline 
Inner radius, $\bar{\xi}_{10}$                                     & 7.7265388(+03)     \\
Inner radius in AU                                                 & 4.5404065(+00)     \\
Outer radius, $\bar{\xi}_{11}$                                     & 1.0659348(+04)     \\
Outer radius in AU                                                 & 6.2638359(+00)     \\
Radius $\alpha_\mathrm{J}$ of $\mathrm{max}|\bar{\theta}|$ in AU   & 5.3334010(+00)     \\
Percentage error in $\alpha_\mathrm{J}$                            & 2.53               \\ 
\hline  
Neptune--Shell No                                                  & 18                 \\
\hline 
Inner radius, $\bar{\xi}_{17}$                                     & 4.6558115(+04)     \\
Inner radius in AU                                                 & 2.7359310(+01)     \\
Outer radius, $\bar{\xi}_{18}$                                     & 5.6506774(+04)     \\
Outer radius in AU                                                 & 3.3205518(+01)     \\
Radius $\alpha_\mathrm{J}$ of $\mathrm{max}|\bar{\theta}|$ in AU   & 3.0148759(+01)     \\
Percentage error in $\alpha_\mathrm{N}$                            & 0.31               \\     
\hline 
\end{tabular}
\end{center}
\end{table}

\subsection{Trans-Neptunian Objects}
\label{TNOs}
Computing quantities related to TNOs constitutes a very-long-distance integration problem, since the corresponding complex IVP is solved along the contour $\mathfrak{C}_{2D}$ (Equation~\eqref{C2D}). A global percentage error $\%\mathcal{E}(\bar{\theta}) \leq 9 \times 10^{-8}$ has been verified for the code \texttt{DCRKF54}, which is quite satisfactory for this case. Our model reproduces the sharp division between ``plutinos'' and ``classical Kuiper belt objects'' (also called ``cubewanos''); in particular, the border between Shell~No~19 and Shell~No~20 is at $\sim 40$~AU, and the ``Kuiper cliff'' appears at $\sim 47$~AU.
  
The TNO Sedna has a highly eccentric orbit with eccentricity 0.853, aphelion 937 AU, perihelion 76.4 AU, semi-major axis 518.6 AU, and inclination 11.9$^o$. Regarding its semi-major axis, it lies inside Shell~No~40 with inner radius $\bar{\xi}_{39} = 478.3 \, \mathrm{AU}$, outer radius $\bar{\xi}_{40} = 521.9 \, \mathrm{AU}$, and radius of $\mathrm{max}|\bar{\theta}|$ $\alpha_{40} = 500.2 \, \mathrm{AU}$ deviating $\sim 3.5\%$ from the observed value. An interesting scenario studied in \cite{GML06} is that: (i) Sedna's orbit has been perturbed by a Jupiter-mass object at 5000 AU. Shell~No~77 has inner radius $\bar{\xi}_{76} = 4888  \, \mathrm{AU}$, outer radius $\bar{\xi}_{77} = 5118 \, \mathrm{AU}$, and radius of $\mathrm{max}|\bar{\theta}|$ $\alpha_{77} = 5001 \, \mathrm{AU}$ deviating $\sim 0.02\%$ from the value computed by this scenario. Alternatively, (ii) Sedna's orbit has been perturbed by a Neptune-mass object at 2000 AU. Shell~No~59 has inner radius $\bar{\xi}_{58} = 1902 \, \mathrm{AU}$, outer radius $\bar{\xi}_{59} = 2020 \, \mathrm{AU}$, and radius of $\mathrm{max}|\bar{\theta}|$ $\alpha_{59} = 1961 \, \mathrm{AU}$ deviating $\sim 2\%$ from the value computed by this alternative scenario. 

The largest TNO Eris has aphelion 97.7 AU, perihelion 38.4 AU, semi-major axis 68.0 AU, eccentricity 0.436, and inclination 43.8$^o$. Concerning its semi-major axis, it lies inside Shell~No~23 with inner radius $\bar{\xi}_{22} = 66.1 \, \mathrm{AU}$, outer radius $\bar{\xi}_{23} = 77.1 \, \mathrm{AU}$, and radius of $\mathrm{max}|\bar{\theta}|$ $\alpha_{23} = 71.4 \, \mathrm{AU}$ deviating $\sim 5\%$ from the observed value. Some remarks made for the second largest TNO Pluto in \cite{GD94} (Section~2, paragraph preceding the last one) do also hold for Eris.

Makemake, the third largest TNO (after Eris and Pluto), has aphelion 53.1 AU, perihelion 38.5 AU, semi-major axis 45.8 AU, eccentricity 0.159, and inclination 28.9$^o$. Regarding its semi-major axis, it lies inside Shell~No~20 with inner radius $\bar{\xi}_{19} = 40.0 \, \mathrm{AU}$, outer radius $\bar{\xi}_{20} = 47.6 \, \mathrm{AU}$, and radius of $\mathrm{max}|\bar{\theta}|$ $\alpha_{20} = 43.6 \, \mathrm{AU}$ deviating $\sim 5\%$ from the observed value. 

Haumea, the fourth largest TNO, has aphelion 51.5 AU, perihelion 34.7 AU, semi-major axis 43.1 AU, eccentricity 0.195, and inclination 28.2$^o$. Its semi-major axis lies inside Shell~No~20 with inner radius $\bar{\xi}_{19} = 40.0 \, \mathrm{AU}$, outer radius $\bar{\xi}_{20} = 47.6 \, \mathrm{AU}$, and radius of $\mathrm{max}|\bar{\theta}|$ $\alpha_{20} = 43.6 \, \mathrm{AU}$ deviating $\sim 1\%$ from the observed value. 

Furthermore, the TNO Quaoar has aphelion 45.1 AU, perihelion 41.7 AU, semi-major axis 43.4 AU, eccentricity 0.039, i.e. the smallest eccentricity among the largest known TNOs, and inclination 8$^o$. Concerning its semi-major axis, it lies inside Shell~No~20 with inner radius $\bar{\xi}_{19} = 40.0 \, \mathrm{AU}$, outer radius $\bar{\xi}_{20} = 47.6 \, \mathrm{AU}$, and radius of $\mathrm{max}|\bar{\theta}|$ $\alpha_{20} = 43.6 \, \mathrm{AU}$ deviating only $\sim 0.5\%$ from the observed value.   

On the other hand, Varuna is the TNO with the second smallest eccentricity, having aphelion 45.3 AU, perihelion 40.5 AU, semi-major axis 42.9 AU, eccentricity 0.056, and inclination 17.2$^o$. Its semi-major axis lies inside Shell~No~20 with inner radius $\bar{\xi}_{19} = 40.0 \, \mathrm{AU}$, outer radius $\bar{\xi}_{20} = 47.6 \, \mathrm{AU}$, and radius of $\mathrm{max}|\bar{\theta}|$ $\alpha_{20} = 43.6 \, \mathrm{AU}$ deviating $\sim 1.5\%$ from the observed value.

It seems that Quaoar and Varuna, due to the very small eccentricities of their orbits, are ideal candidates for obeying the global polytropic model (for similar comments regarding jovian satellites, see \citep{GV94}, Section~2; for comments regarding planets, see \citep{GD94}, Section~2)). In fact, an object in almost circular orbit implies that it has evolved under mild processes, which, in turn, maintain the sensitive global polytropic character of such a system.

\section{Remarks and Conclusion}    
\label{remarks}         
First, it is worth emphasizing that in this study the solar and jovian systems are considered within the framework of classical mechanics. In particular, it is assumed that these systems obey globally the equations of hydrostatic equilibrium. We mention here that the simulations of several astrophysical systems by polytropic models is a well-established long-lived hypothesis in astrophysics; readers interested in polytropic models can find full details and plethora of astrophysical applications in \citep{Hor04} (for the solar system, see Section~6.1.7). A second interesting remark is that there is in fact only one parameter, which must be adjusted: the polytropic index $n$ of the central body. The algorithm for computing an optimum value $n_\mathrm{opt}$ is described and used in \citep{GV94} for the jovian system (as discussed in Section~\ref{polSJS}) and in \citep{GV93} for the solar system. Third, it is worth emphasizing that there are not any ``external parameters'' (i.e. empirical, semi-empirical, borrowed from other theories, etc.) inserted into our simulations. The quantized orbits are emerging as intrinsic properties of the system under consideration. In particular, each quantized orbit is identified and computed as the distance of the local density--extremum of each polytropic shell (belonging to the resultant polytropic configuration) from the central body. In the numerical treatment of the problem, the Fortran code DCRKF54 has been used, which can solve complex IVPs along complex paths. So, as said in \citep{Ger93} (Abstract), the theoretical input to the global polytropic model is very simple, while the numerical output admits of several interesting physical interpretations.   

Regarding alternative studies on quantized orbits of planets and satellites, it is first worth remarking that there is a common practice in the majority of these studies to use several external parameters. Such parameters have mainly to do with the well-known ``Titius-Bode (TB) law'' or with several modifications of this law, or with several TB-type laws. We mention here that the TB law is an empirical formula involving some parameters, which are fixed by observation(s). In view of certain assumptions, the TB law can be written so that to resemble the ``quantized Bohr atomic model'' (see e.g. \citep{PPL08}, Section~1; \citep{G07}, Section~1; \citep{AF97}, Section~2). Accordingly, quantum mechanics enters into the scene of such considerations (for a discussion on the similarities with quantum mechanics as well as on the uncertainty of such approach, see e.g. \citep{AF97}, Sections~3--5). In this case the ``Bohr radius'' of a planetary or a satellite system seems to be of great importance (\citep{PPL08}, Equation~(5); \citep{AF97}, Equation~(15); \citep{RR98}, Equation~(7)). A relevant  quantum-like approach is to set up some Schr\"{o}dinger-type equations (see e.g. \citep{OMC04} and references therein).
Finally, there are alternative studies using: (i) Scale Relativity (\citep{HSG98} and references therein), which is an extension of Einstein's principle of relativity: by giving up the differentiability of space-time coordinates at very large time-scale, the solar system can be described by a Schr\"{o}dinger-type equation; and (ii) post-Newtonian approximations due to the finite propagation speed of gravitational interaction (\citep{G07} and references therein).  

Concluding, we emphasize on the fact that several predictions made recently by the above alternative studies, can be also found in the numerical results of the global polytropic model. We mention indicatively some such results. First, two ``intramercurial orbits'' with radii $\sim 0.05$ and $\sim 0.18$ AU, which are empty in our solar system but they are observed to be occupied in several extra-solar planetary systems (\citep{OMC04}, Section~1), are quoted in Table~III of \citep{Ger93}, where the unoccupied Shells No~3 and No~4 are shown with radii of $\mathrm{max}|\bar{\theta}|$ $\alpha_{3} \sim 0.05 \, \mathrm{AU}$ and $\alpha_{4} \sim 0.15 \, \mathrm{AU}$, respectively. Second, the radius of the Neptune's orbit, which is inaccurately estimated in several studies, is computed by the global polytropic model with a satisfactory accuracy (Table~\ref{sun}, results in group~4). Third, the recently discovered asteroids orbiting between Uranus and Neptune with orbit radii $\sim 24.8$ AU (\citep{OMC04}, Section~1), seem to occupy Shell No~17 as quoted in Table~III of \citep{Ger93} with radius of $\mathrm{max}|\bar{\theta}|$ $\alpha_{17} \sim 24.7 \, \mathrm{AU}$. Finally, the radius of the Quaoar's orbit (\citep{OMC04}, Section~4) has been computed in the present work with a satisfactory accuracy (Section~\ref{TNOs}).
Readers interested in the issues of this study can find further details in \citep{Ger93}, \citep{GV94}, \citep{GD94}, and \citep{GV93}.

\end{document}